\documentclass[prl,preprint,showpacs,preprintnumbers,amsmath,amssymb]{revtex4}
\usepackage{graphicx}
\usepackage{subfigure}
\usepackage[citecolor=blue,colorlinks=true,linkcolor=blue]{hyperref}
\usepackage{dcolumn}

\begin{document}

\title{Island Size Selectivity during 2D Ag Island Coarsening on Ag (111) }

\author{Giridhar Nandipati}
\email{gnandipa@mail.ucf.edu	}
\author{Abdelkader Kara}
\email{akara@mail.ucf.edu}
\author{Syed IslamuddinShah}
\author{Talat S. Rahman}
\email{trahman@mail.ucf.edu}
\affiliation{Department of Physics, University of Central Florida,  Orlando, FL  32816}

\date{\today}

\begin{abstract}
We  report on early stages of submonolayer Ag island coarsening on Ag(111) surface at room temperature ($300$ K)  carried out using realistic kinetic Monte Carlo (KMC) simulations.
We find that during early stages, coarsening proceeds as a sequence of selected island sizes creating peaks and valleys in the island size distribution.
We find that island-size selectivity is due to formation of kinetically stable islands for certain sizes because of adatom detachment/attachment processes and large activation barrier for kink detachment. 
 In addition, we find that the ratio of number of adatom attachment to detachment processes to be independent of parameters of initial configuration and also on the initial shapes of the islands confirming that island-size selectivity is independent of initial conditions.These simulations were carried out using a very large database of  processes identified by their local environment and whose activation barriers  were calculated using the embedded-atom method.

\end{abstract}
\maketitle
The phenomenon of coarsening or ripening plays an important role in a wide variety of processes in many branches of the physical sciences. Of particular technological interest is coarsening of two or three-dimensional islands on a surface \cite{Zinke}, since coarsening determines the nanoscale ordering and surface structure. As a result island coarsening has recently been the subject of a great deal of experimental and theoretical investigation. \cite{Zinke, pai1,Khare, Meakin, Soler, sholl, Shollprl,Kandel, Zinke2, pslkmc, pkmc, OstwaldAg}. Ostwald ripening \cite{ostwald, voorhees, LS} is a general feature at late stages of phase separation and can be used for island decay on metal and semiconductor surfaces \cite{theis1,theis2,morgenstern1}.
During Ostwald ripening (OR), driven by lowering of excess surface free energy associated with island edges, islands larger than a critical size grow at the expense of smaller islands.  This results in a power-law growth of the average islands size with an exponent of $1/3$ if the adatom diffusion is rate limiting processes. In OR, islands are assumed to be immobile and coarsening is mediated by diffusion of atoms between islands.
Scanning tunneling microscopy (STM) studies have revealed that OR dominates the coarsening of island distributions for Ag/Ag(111)\cite{morgenstern1,morgenstern2} at room temperature.

In this article, we  present results of our study of initial stages of coarsening of Ag islands on Ag(111) surface using kinetic Monte Carlo (KMC) simulations at room temperature ($300$ K). Our realistic KMC simulations were carried out using a very large database of processes with corresponding energy barriers based on the local neighborhood. We studied how island size distribution (ISD) changes as coarsening proceeds depending on the starting ISD. Although, most of the results shown are for Gaussian initial ISD,  we have also carried out simulations starting with delta ISD. In addition, we have studied the effect of island shapes in the starting ISD on initial stages of coarsening.

The database of processes was obtained from previous self-learning KMC (SLKMC) \cite{slkmc1,slkmc2,offkmc} simulations of small and large Ag island diffusion on Ag(111) surface, carried out at $300$ and $500$ K .
 In determining all activation energies we used  interaction potentials based on the embedded-atom method (EAM) as developed by Foiles {\it et al} \cite{Foiles}. In an SLKMC simulation, rather than using a fixed catalog of processes and their corresponding activation barriers they are obtained on the fly whenever a new configuration is found that is not already included in the database. Our KMC simulations in this work are SLKMC simulations with a closed database.
In particular, we have used this database to do longer timescale KMC simulations of Ag(111) island coarsening at room temperature. 
We used the same prefactor for both single and multiatom diffusion processes. It has been shown that prefactor is same for adatom diffusion processes \cite{handan} but no  detailed investigations for multiatom diffusion is available.  
More details about how this database was obtained and  modifications and improvements for speeding up KMC simulations can be found in ref.\onlinecite{pslkmc}.
 
The initial configuration for these coarsening simulations was created by dividing the empty lattice into boxes and placing islands of different sizes randomly at the center of these boxes to prevent overlap of islands.
The number of islands of a particular size depends on whether the starting ISD is a Gaussian or a delta function. To avoid finite-size effects we carried out our simulations using a relatively large system size $L=1024$ with periodic boundary conditions and in order to obtain good statistics we averaged our results over $10$ runs.

Our starting ISD with a Gaussian distribution has a total of $742$ islands with a peak of $100$ islands at the average island size and  a width of $3$.
The total number of islands in the starting Gaussian ISD with a different average island size remains the same as long as the peak island count and the width of the  distribution are kept constant.
We note that the total number of islands in the delta ISD is chosen to be same as in a Gaussian ISD. 
For simplicity, the initial 
shapes of islands were chosen arbitrarily and islands of same size have same shape. Most of these shapes are either compact or close to compact. 
 
 \begin{figure}
\subfigure{ \includegraphics [width=5.5cm]{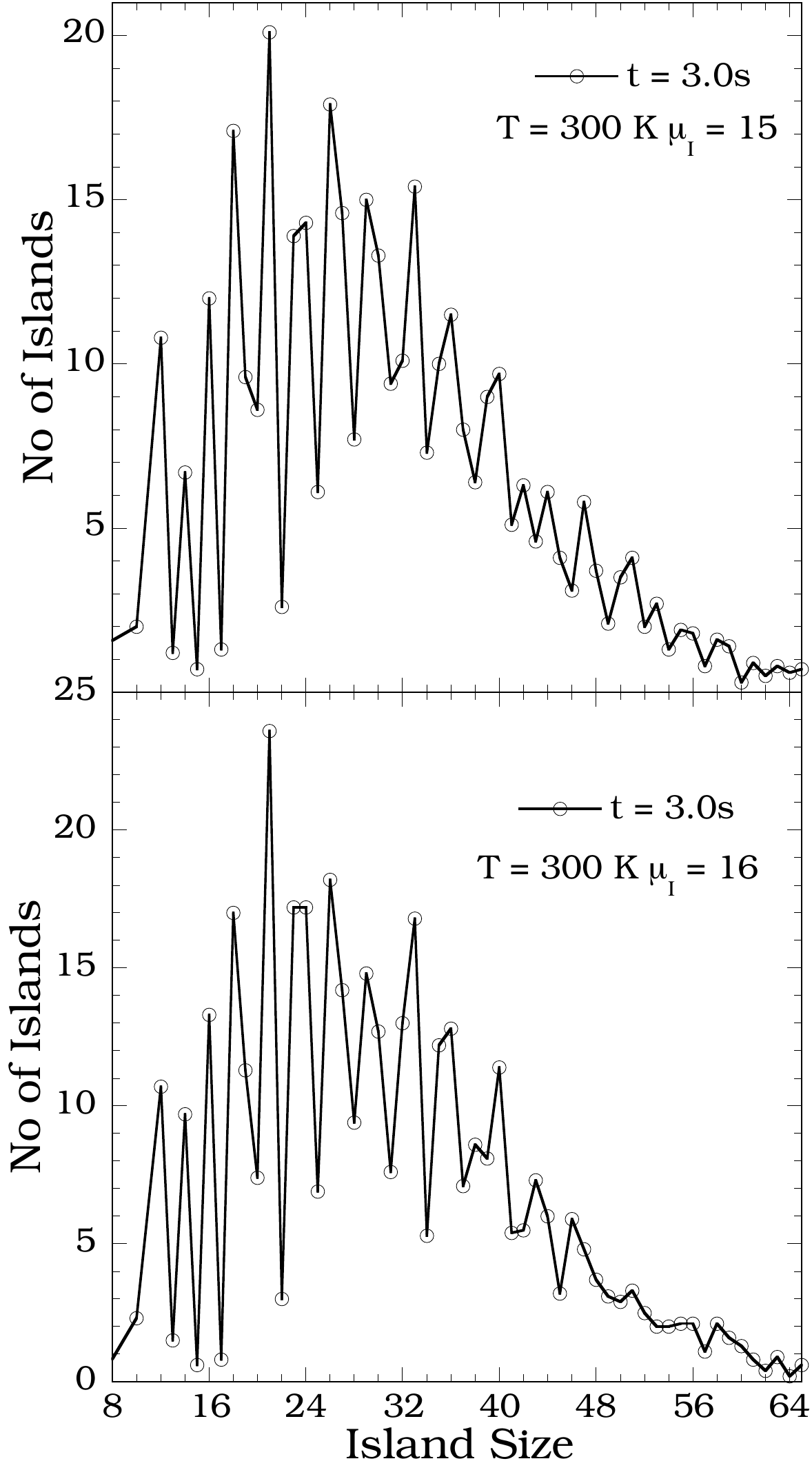} }
\caption{\label{isd_300k_3s}{Island size distributions  after $3$ s when the initial size distribution is a Gaussian for average island sizes ($\mu_{I}$) $15$ and $16$ at T = $300$ K.}}

\vspace{10pt}
\subfigure {\includegraphics [width=5.5cm]{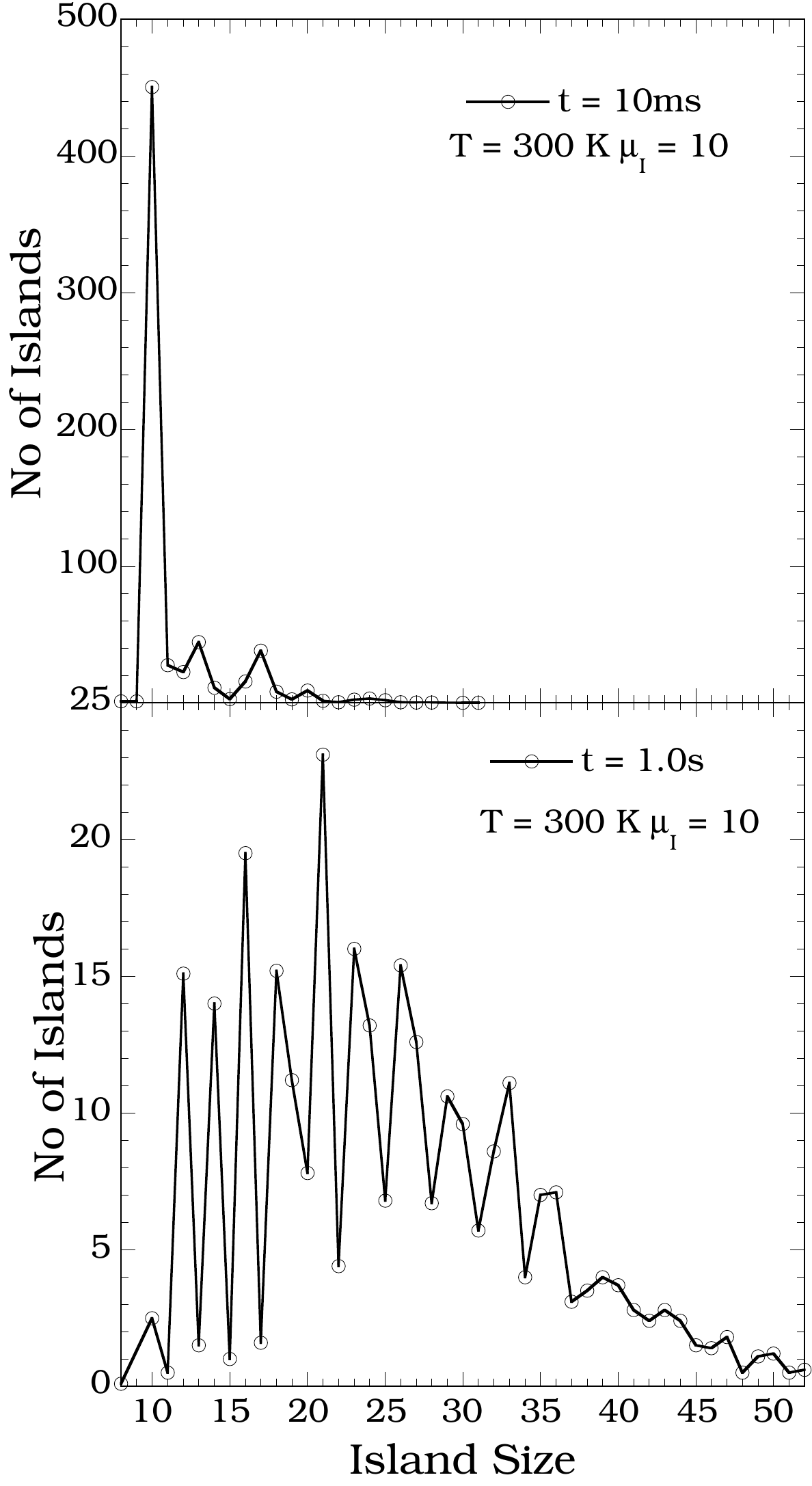} }
\caption{\label{isd_delta}{Island size distributions  $10.0$ ms and $1.0$ s when the initial size distribution is a delta function for average island size ($\mu_{I}$) $10$ at T  = $300$ K}}

\end{figure}

To study coarsening behavior we looked at island size distributions at different times as coarsening proceeds for different initial configurations.  To study early stages of coarsening we carried out simulations only up to $3$ seconds.
Fig.~\ref{isd_300k_3s} shows island size distributions after $3$ s of coarsening for initial average island sizes $15$ and $16$ when the initial ISD is a Gaussian, while Fig.~\ref{isd_delta} shows island size distributions after $10$ms and $1$ s of coarsening  for initial average island size $10$ when the initial ISD is a delta function. As can be seen, during coarsening there is a dramatic change in the island size distribution from a smooth Gaussian distribution or delta distribution to a non-smooth distribution with peaks and valleys at specific island sizes.  
From Figs.~1 \& 2 it can be seen that island sizes whose populations are either a peak or valley in the ISD do not change as the coarsening proceeds.
The same behavior of ISD with peaks and valleys was observed even for other parameters of initial configuration and appears to be independent of type  of initial ISD and the initial averages island size.
Table~\ref{table1} summarizes island sizes up to  $35$ according to whether they constitute a peak, valley or neither of the ISD  after $1.0$ s of coarsening.

\begin{table}
\caption{\label{table1} List of islands size where the ISD is a peak, valley or neither after $1.0$ s coarsening.}
 \begin{tabular}{ c c c}
\hline
Valley & Peak & neither\\
\hline
  11 & 12  \\
  13 & 14  \\
  15 & 16  \\
  17 & 18 & 19\\
  20 & 21 \\
  22 & 23(24)\\
  25 & 26  & 27\\
  28 & 29  &30 \\
  31 & 33  & 32\\
  34 & 35\\
  \hline
\end{tabular}
\end{table}
 
\begin{figure}[h]
\includegraphics [width=6.5cm]{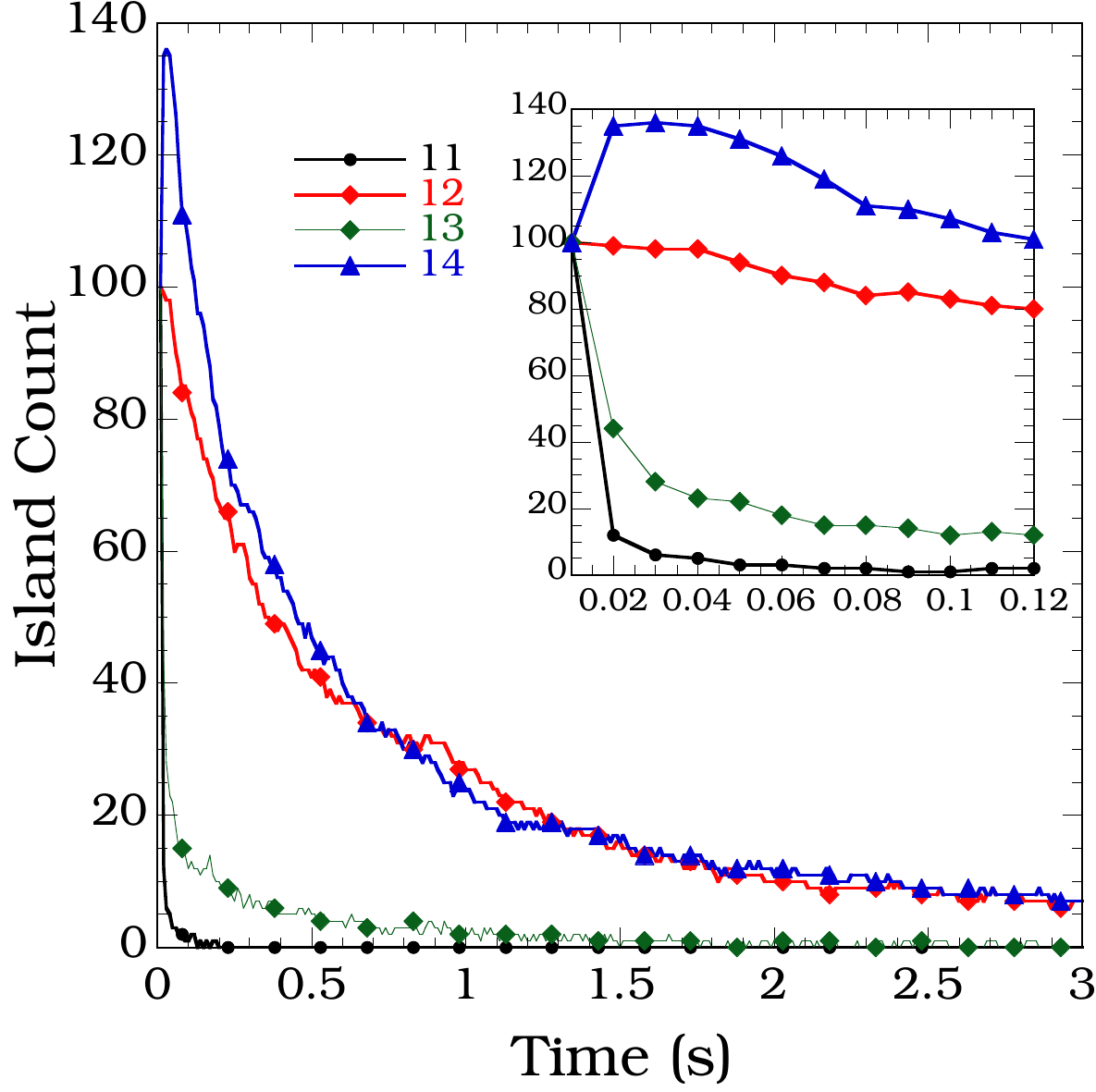} 
\caption{\label{Idecay}{Decay of island density for different  island sizes at T = $300$ K. At $t=0$ peak island count is $100$}}
\end{figure}

Since the results for delta ISD (Fig.~\ref{isd_delta}) were the same,  in what follows we concentrate our discussion on the results for Gaussian initial ISD.
 Fig.~\ref{Idecay} shows decay of number of islands for different sizes with time for a Gaussian initial ISD.  Note that the number of islands for sizes $11$ and $13$  are valleys while those for sizes $12$ and $14$  are peaks in the ISD.
 From fig.~\ref{Idecay}, it can be seen that the number of islands of size $11$ and $13$ (valleys in the ISD) decay exponentially in the very first few microseconds of coarsening. For islands of size $12$ and $14$ (peaks in the ISD), their densities increase for the first few microseconds before starting to decay at a much slower rate. The same pattern emerges for all island sizes constituting peaks or valleys.
This suggests that island sizes whose population are peaks in the ISD form kinetically stable islands.
 
 Peaks and valleys in the ISD during coarsening and differences in the rate of decay of densities of corresponding island sizes suggest that coarsening occurs though a sequence of selected island sizes.   Similar decay of island densities is also  observed  when the initial ISD  is a delta function (see fig.~\ref{isd_delta}). We note also that for some island sizes ($19$, $27$ and $30$) the ISD has neither a peak nor a valley (see table~\ref{table1}). For island sizes $23$ and $24$, either one can be a peak, but for the most part, $23$ is a peak, while $24$ is neither a peak nor a valley. 

\begin{figure}[h]
\includegraphics [width=5.4cm]{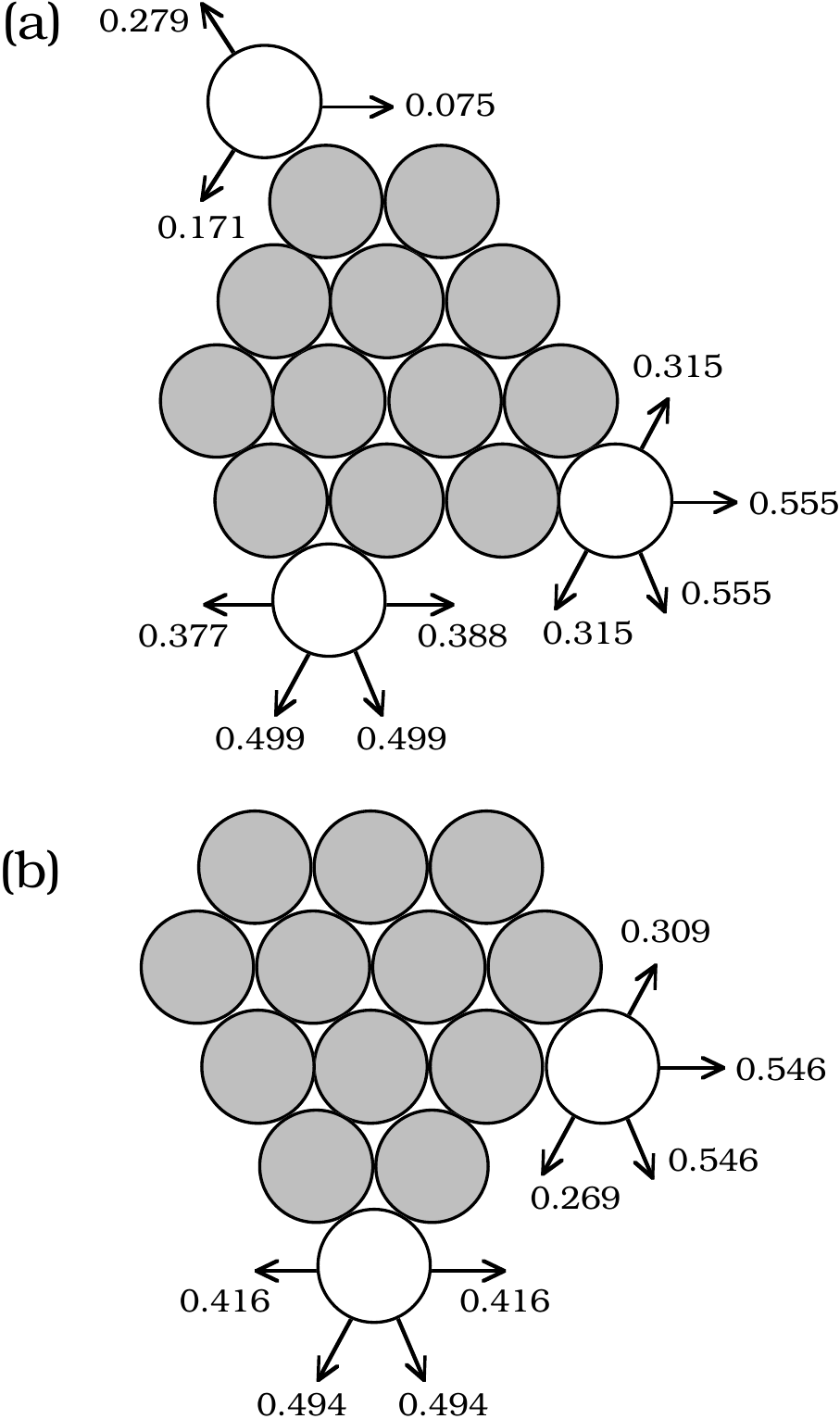} 
\caption{\label{detach_processes}{Activation barriers (in eV) for the most frequent detachment processes and for edge diffusion processes}}
\end{figure}

From experimental\cite{morgenstern1, OstwaldAg} and theoretical studies \cite{pslkmc} it is known that for 2D Ag/Ag(111) coarsening is due to evaporation-condensation mediated by monomer diffusion between islands. In order to understand island-size selectivity, we looked at detachment processes on the basis of island size. We find that for all island sizes  larger $8$ the most frequent detachment process is an atom's detaching from a step edge to create a monomer.
In addition, we find that the number of events of edge atom detachment for island sizes whose populations are valleys in the ISD is higher than for island sizes whose populations are peaks in the ISD.
Fig.~\ref{detach_processes} shows the most frequent detachment processes (detachment of edge atom) for islands along with their corresponding activation barriers while the energy barrier for atoms  with at least $3$ nearest neighbors to detach is greater than $0.7$ eV and they rarely detach to create monomers at room temperature
This indicates that islands of different sizes whose population are valleys in the ISD, that are formed during coarsening usually have an edge atom in their shapes. 
Any island with an edge atom either loses it because of detachment leaving a smaller island of selected size or less frequently attracts nearby monomer creating a bigger island of selected size, as a result creating island sizes whose population are peaks in the ISD and valleys at other island sizes. Island sizes whose populations are peaks in the ISD do not have any edge atoms: all atoms have at least $3$ nearest neighbors making them kinetically stable islands.
We observed the same behavior for the  ISD as coarsening proceeds for all types of initial island size distributions and initial average island sizes. In addition, ISD has the same behavior if the shapes of islands were changed in the initial configuration. In particular, we observed the same behavior when started with an initial configuration in which all islands are either have a kinetically stable shape or low energy shape. 
This suggests that island-size selectivity is independent of parameters of initial ISD and also on shapes of the islands in the initial configuration.

 \begin{figure}[h]
\includegraphics [width=6.5cm]{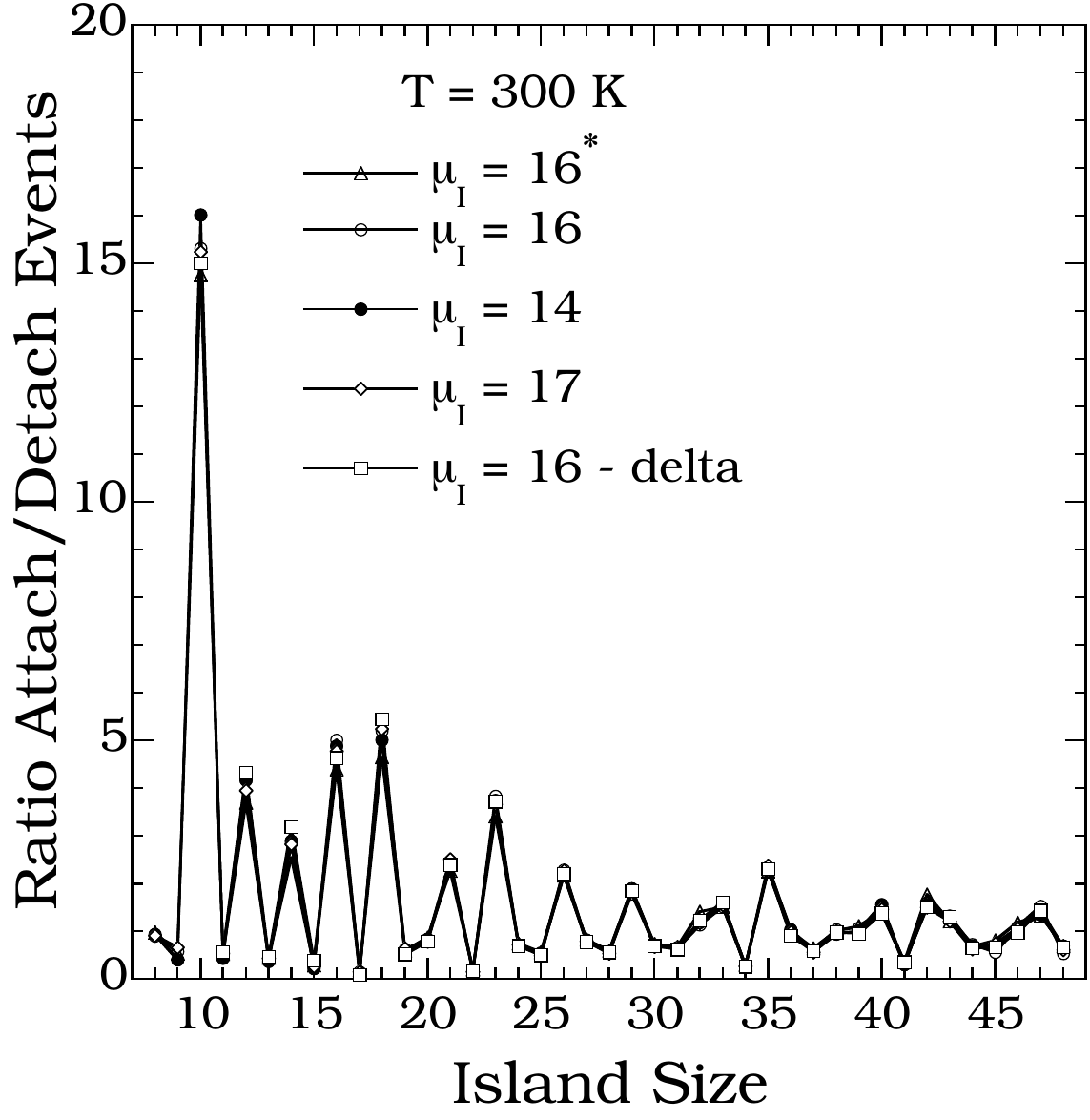} 
\caption{\label{ad_ratio_all}{Ratio of number of attachment to detachment events at $300$ K after $3$s of coarsening for
Gaussian initial ISD ($\mu_{I} = 14, 16,17$), delta initial ISD ($\mu_{I} = 16$) and Gaussian initial ISD with islands either having a kinetically stable or low enegy shapes ( $\mu_{I} = 16^{*}$) }}
\end{figure}

Fig.~\ref{ad_ratio_all} shows the ratio of number of attachment to detachment events at $300$ K after $3$ s of coarsening when started with different initial average island sizes for a Gaussian initial ISD and delta initial ISD. 
Fig~\ref{ad_ratio_all} also shows the ratio for average island size $\mu_{I} = 16^{*}$ when the coarsening is started with Gaussian initial ISD and in which all islands have shapes that are either low energy or kinetically stable. 
It can be seen that peaks and valleys are exactly at the same island size as in the island size distributions. This suggest that densities of selected island sizes decay because of attachment events while non-selected island sizes decay because of detachment of edge atoms. Surprisingly, they all collapse to one single curve, indicating that this ratio is independent of all parameters for initial ISD and also of the shapes of the islands in the initial configuration.  From this we may conclude that island size selectivity is the behavior of the early stages of Ag/Ag(111) coarsening and is independent of initial configuration used to start the coarsening.

In summary, we find that during early stages,  2D Ag/Ag(111)  island coarsening proceeds as a sequence of selected island sizes. 
ISD quickly changes to a non-smooth distribution with peaks and valleys owing to differences in the rate of decay of selected and non-selected island sizes. Densities of non-selected island sizes decay exponentially in the first few microseconds of coarsening.
This decay of island densities of non-selected island sizes  is primarily due to detachment of edge atoms, creating monomers and smaller islands of selected size. 
Densities of selected island sizes, on the other hand decay at a much slower rate because of the formation of kinetically stable shapes. 
 In an island of kinetically stable shape, all atoms have at least $3$ nearest neighbors and because of high detachment barrier they rarely detach from the island to create monomers at room temperature.
 Accordingly, we conclude that island size selection is primarily due to adatom  detachment/attachment processes at island boundaries and the higher energy barrier to detachment for atoms with at least $3$ nearest neighbors.  
 Elsewhere we show that these factors also restrict the shapes of island that can form during coarsening\cite{iss_long}.
In particular, the share of  kinetically stable islands formed during coarsening is greater for selected island sizes than for non-selected island sizes.
Moreover, certain non-selected island sizes, though a kinetically stable shape might exist, appear only rarely, since except through detachment atoms never rearrange themselves into a kinetically stable shape.
Recall from Figs.~\ref{isd_300k_3s} \& \ref{isd_delta} that the densities of certain non-selected island sizes ($11$, $13$, $15$, $17$) are either zero or very close to zero, while others ($20$, $22$, $25$, $28$, $31$, $34$)are not. These later sizes possess kinetically stable shapes \cite{iss_long}.
Finally, the collapse of the ratio of the number of attachment to detachment events into a single curve (see fig.~\ref{ad_ratio_all}) for different parameters of initial configuration suggests that island-size selectivity is independent of all of those.
 
This work was supported by NSF grant ITR-0840389. We would also like to acknowledge grants of computer time from the Ohio Supercomputer Center (grant no. PJS0245) and also computational resources provided by University of Central Florida. We thank Lyman Baker for critical reading of the manuscript.

\bibliography{references}

 \end{document}